\documentclass[12pt]{article}
\usepackage[dvips]{graphicx}
\usepackage[latin1]{inputenc}
\def\be{\begin{equation}}
\def\ee{\end{equation}}
\def\ba{\begin{eqnarray}}
\def\ea{\end{eqnarray}}

\newcommand{\II}{\mbox{\tiny I}}
\newcommand{\XY}{\mbox{\tiny XY}}

\begin{document}
\begin{titlepage}
\thispagestyle{empty}
\vskip0.5cm
\begin{flushright}
MS--TPI--98--13
\end{flushright}
\vskip1.5cm

\begin{center}
{\Large {\bf Monte Carlo Algorithms}}
\vskip3mm
{\Large {\bf For the Fully Frustrated XY Model}}
\end{center}

\vskip1.0cm
\begin{center}
{\large S. Gro\ss e Pawig and K. Pinn}\\
\vskip5mm
Institut f\"ur Theoretische Physik I, Universit\"at M\"unster\\ 
Wilhelm--Klemm--Str.~9, D--48149 M\"unster, Germany \\[2mm]
e--mail: {\sl pawig@math.uni--muenster.de, pinn@uni--muenster.de}
\end{center}

\vskip2.5cm
\begin{abstract}
\par\noindent
We investigate local update algorithms for the fully frustrated XY
model on a square lattice.  In addition to the standard updating
procedures like the Metropolis or heat bath algorithm we include
overrelaxation sweeps, implemented through single spin updates that
preserve the energy of the configuration. The dynamical critical
exponent (of order two) stays more or less unchanged.
However, the integrated
autocorrelation times of the algorithm can be significantly reduced.
\end{abstract}
\end{titlepage}

\section{Introduction}
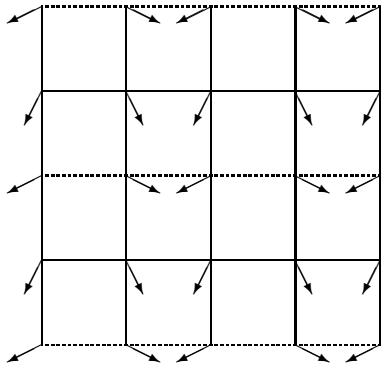
\begin{figure}
\unitlength0.75mm
\begin{center}
\begin{picture}(60,65)
\put( 0,5){\dashbox{0.5}(60,30){}}
\put( 0,50){\dashbox{0.5}(60,15){}}
\put( 0,5){\line(0,1){60}}
\put(15,5){\line(0,1){60}}
\put(30,5){\line(0,1){60}}
\put(45,5){\line(0,1){60}}
\put(60,5){\line(0,1){60}}
\put( 0,20){\line(1,0){60}}
\put( 0,50){\line(1,0){60}}
\put( 0,5){\vector(-2,-1){6}}
\put(15,5){\vector( 2,-1){6}}
\put(30,5){\vector(-2,-1){6}}
\put(45,5){\vector( 2,-1){6}}
\put(60,5){\vector(-2,-1){6}}
\put( 0,20){\vector(-1,-2){3}}
\put(15,20){\vector( 1,-2){3}}
\put(30,20){\vector(-1,-2){3}}
\put(45,20){\vector( 1,-2){3}}
\put(60,20){\vector(-1,-2){3}}
\put( 0,35){\vector(-2,-1){6}}
\put(15,35){\vector( 2,-1){6}}
\put(30,35){\vector(-2,-1){6}}
\put(45,35){\vector( 2,-1){6}}
\put(60,35){\vector(-2,-1){6}}
\put( 0,50){\vector(-1,-2){3}}
\put(15,50){\vector( 1,-2){3}}
\put(30,50){\vector(-1,-2){3}}
\put(45,50){\vector( 1,-2){3}}
\put(60,50){\vector(-1,-2){3}}
\put( 0,65){\vector(-2,-1){6}}
\put(15,65){\vector( 2,-1){6}}
\put(30,65){\vector(-2,-1){6}}
\put(45,65){\vector( 2,-1){6}}
\put(60,65){\vector(-2,-1){6}}
\end{picture}
\parbox[t]{.85\textwidth}
 {
 \caption[ZQX]
 {\label{figGS}
\small
A ground state configuration of the 
FFXY model. The dotted lines represent links with antiferromagnetic
couplings, the full lines are links with ferromagnetic couplings.
}}
\end{center}
\end{figure}

The 2-dimensional fully frustrated XY (FFXY) model has attracted a lot
of attention over the years. It is defined through the partition
function 
\be Z = \int \prod_{i} \left( d^2 \! s_i \, \delta \left(\vec{s_i}^2 -
  1\right) \right) \exp\left[\beta \sum_{<ij>} J_{ij} \, \vec{s}_i \cdot 
  \vec{s}_j\right] \, .  
\ee 
The 2-component spins $s_i$ are defined on a square grid. They are
constrained to unit length, $\vec{s_i}^2= 1$.  $\beta > 0$ denotes the
inverse temperature.  Units are chosen such that $k_B \equiv 1$. The
$J_{ij}$ are either $+1$ or $-1$ in such a way that each elementary
plaquette of the lattice contains exactly one antiferromagnetic bond
(with $J=-1$). We stick to the convention that the
antiferromagnetic links are those lying in every second row of the
lattice, see figure~\ref{figGS}, which shows also one of the ground
states of the FFXY model.

The importance of the frustrated XY model stems from the fact that it
provides a convenient framework to study a variety of interesting
phenomena displayed by numerous physical systems. Experimental
realizations of the model are, e.g., 2-dimensional arrays of Josephson
junctions or superconducting wire networks
\cite{review,denniston,ling}.

The phase structure of the FFXY model is still debated.  In the
literature, there seems to be agreement about the following: In
addition to the type of excitations present in the unfrustrated XY
model \cite{KT} (spin waves at large $\beta$ and vortices for
$\beta \leq \beta_{KT}$) there is a new type of 
discrete excitations which should be relevant at sufficiently large
$\beta$. With each elementary plaquette $i$ one can associate a chirality
variable $\sigma_i$, defined through
\begin{eqnarray}
  \sigma_i &=& 
{\rm sign}\left(\sin\phi_{12} + \sin\phi_{23}
                         + \sin\phi_{34} + \sin\phi_{41} \right) \, , 
              \label{vort} \\
           \phi_{kl} &=& \theta_k - \theta_l - A_{kl} \, . \nonumber
\end{eqnarray}
The corners of the plaquette $i$ are labelled clockwise $1 \dots 4$. 
The angles $\theta_k$ are defined through the relation $\vec{s}_k=
(\cos\theta_k,\sin\theta_k)$, and $A_{kl}= 0$ or $\pi$ for ferromagnetic
and antiferromagnetic links, respectively.  $\sigma_i$ can assume
values $\pm 1$.  The 
global symmetry of the model with respect to the reflection of all
$\sigma$'s can be spontaneously broken. If this transition takes place
at a $\beta$-value different from that where the XY degrees
of freedom undergo a (most likely) Kosterlitz-Thouless phase
transition, then it follows from standard wisdom about universality
that it should belong to the universality class of the 2-dimensional
Ising model.  On the other hand, if the two transitions are exactly on
top of each other, one expects a new universality class, with
exponents different from the 2D Ising values.

The critical properties of the FFXY model have been investigated in a
number of Monte Carlo studies, see, e.g.,
refs.~\cite{Teitel}--\cite{Olsson2}.  All studies, including
the most recent ones, indicate that there is an urgent need to go to
larger lattices. Unfortunately, the simulations of the FFXY model
exhibit strong critical slowing down. On large lattices one has
to perform many update sweeps in order to obtain a
statistically independent configuration. E.g., at $\beta= 2.2$, which
is in the critical region, the autocorrelation time of a
local Metropolis algorithm is $\approx$ 1600 on a 64 by 64 lattice.
Fortunately, there seems to be no exponential slowing down.
At criticality the
autocorrelations grow like $\tau \simeq A \, L^z$, where $L$
denotes the linear lattice extension. The dynamical critical exponent
is about 2.3 for  a local Metropolis update algorithm. 

It is tempting to apply cluster algorithms \cite{SW,Wolff} to the
problem. However, as has been experienced quite often in the case of
frustrated models, there is no obvious way to avoid cluster size
distributions with strong weight on very large cluster sizes, thus
spoiling any gain in efficiency.  Using the general formalism of
Kandel et al.~\cite{KanDom}, with the useful extensions of
ref.~\cite{Coddington}, one can convince oneself that (distinct from
the case of the fully frustrated Ising model \cite{KBD}) in a quite
large class of algorithms there is none that avoids this problem.

In ref.~\cite{Scheinine} an attempt was made to improve on
the performance of the usual local heat bath or Metropolis algorithm,
employing Fourier acceleration. 
However, there was
negligible speedup of observables built
from the discrete $\sigma$-variables introduced above. 
In ref.~\cite{Nicolaides} a Hybrid Monte Carlo algorithm was used, but
without a quantification of critical slowing down.

We here investigate the inclusion of microcanonical
overrelaxation steps \cite{Adler}-\cite{HorKen}
in the update procedure. To the best of our knowledge
it has not yet been employed in the context of the 
frustrated XY model.
The algorithm
is very easy to implement, keeps the updating perfectly local and thus
allows simple vectorization and/or parallelization. Still,
an effective speedup of one order of magnitude can be obtained.

\section{Observables and Algorithms}

In order to test a number of algorithms, we measured for both 
the XY degrees of freedom $\vec{s}$ and the Ising type 
variables $\sigma$ the energies, 
\ba
E_{\XY} &=& \langle \vec{s_i} \cdot \vec{s_j} \rangle \, , \nonumber \\ 
E_{\II}  &=& \langle \mu_i \, \mu_j  \rangle  \, . 
\ea
Here, $i$ and $j$ are nearest neighbours in the lattice, and 
the variables $\mu_i$ are defined through 
\be
\mu_i = (-1)^{i_x + i_y} \, \sigma_i \, , 
\ee
where the lattice coordinate $i$ has components $(i_x,i_y)$.
Furthermore, we measured the susceptibilities 
\ba 
\chi_{\XY} &=& \sum_{j} \langle \vec{s_i} \cdot \vec{s_j} \rangle \, ,
 \nonumber \\
\chi_{\II}  &=& \sum_{j} \langle \mu_i \, \mu_j  \rangle    \, .
\ea 
$i$ is any lattice site, and $j$ runs over the whole grid. 
All our simulations were done on $L$ by $L$ lattices, $L$ even, 
with periodic boundary conditions. 

In order to quantify the critical slowing down we measured the 
integrated autocorrelation times, defined through 
\be 
\tau_{{\rm int},X} = \frac12 \sum_{t=0}^{\infty} 
C_{XX}(t) \, , 
\ee 
using the selfconsistent window method proposed in~\cite{Sokal}.
$C_{XX}$ denotes the normalized autocorrelation function of 
the observable $X$. 

Table \ref{heatL8} shows the integrated correlation times of the four
observables defined above, for $L=8$. Here and in the following 
we chose $\beta=2.2$. This value is compatible with most
estimates of the critical inverse temperature in the literature.
Our results
for $\tau_{\rm int}$ were obtained from $10^6$ lattice sweeps with a
heat bath algorithm.  $\chi_I$ has the longest
autocorrelations. This remains so also on other lattice sizes and also
with any other algorithm that we investigated.

\begin{table}
\small
\begin{center}
  \begin{tabular}{|c|r|}
 \hline
obs  & $\tau_{\rm int}$ \phantom{xx} \\
\hline 
$E_{\II}$              &   9.96(16) \\
$\chi_{\II}$           & 11.16(20) \\
$E_{\XY}$       & 8.36(12)  \\
$\chi_{\XY}$    & 8.12(12)  \\         
\hline
  \end{tabular}
\parbox[t]{.85\textwidth}
 {
\caption{\label{heatL8}
\small
Integrated autocorrelation times 
on an $8\times 8$ lattice, heat bath algorithm, $\beta=2.2$.
}}
\end{center}
\end{table}

In the following, we shall report on results obtained with 
four types of algorithms. In all cases the sweeps are 
done in a typewriter fashion.

\vskip3mm
\noindent {\bf Metropolis (M)}: \\
At site $i$, a rotation of the spin 
by an angle $\theta$ is proposed, where $\theta$ is chosen 
with uniform distribution from the interval $[-\epsilon,+\epsilon]$.
$\epsilon$ can be used to tune the acceptance rate. 
A Metropolis hit is accepted with probability 
${\rm min}[1,\exp(-\beta \Delta E)]$, where $\Delta E$ denotes the 
change of energy associated with the proposed rotation.
At site $i$, one performs $n_{\rm hit}$ Metropolis hits. 

\vskip3mm
\noindent {\bf Metropolis Reflection (MR)}: \\
This is a Metropolis algorithm based on the operation which lies at the 
heart of the Wolff cluster algorithm for the XY model. A single 
random unit vector $\vec{u}$ is selected. One sweeps through the lattice. 
At site $i$, the spin $s_i$ is decomposed in components parallel 
and perpendicular to $\vec{u}$. It is proposed to change the sign 
of the parallel component.
The reflection is actually performed with a probability determined 
by the Metropolis rule.

\vskip3mm
\noindent {\bf Heat Bath (HB)}: \\
We employ a heat bath routine for the generation of 
U(1) random numbers developed by 
Hattori and Nakajima \cite{hattori}, based on a rejection
method with transformation of variables. It has a high 
acceptance rate for all values of the temperature. 

\vskip3mm
\noindent {\bf Microcanonical Overrelaxation (OR)}: \\
Single spins are moved such that the energy remains unchanged. The 
contribution to the total energy of the system that depends on 
a spin $\vec{s_i}$ is proportional to $\vec{s_i}\cdot \vec{h_i}$, 
where $\vec{h_i}$ is a weighted sum of the nearest neighbour spins 
of $\vec{s_i}$. The spin $\vec{s_i}$ is decomposed into components 
parallel and perpendicular to $\vec{h_i}$. The latter component
is reflected (multiplied by $-1$) with probability 1, leaving
unchanged the scalar product $\vec{s_i}\cdot \vec{h_i}$.
Note that the OR algorithm is non-ergodic. It can only be used in a
mixture with an ergodic algorithm. 

\section{Metropolis and Heat Bath Algorithms}

We started our investigation with a study of the first three
algorithms to compare their efficiency.  In table~\ref{compL8} we
quote autocorrelation times for the ``slowest'' observable
$\chi_{\II}$ for the Metropolis-Algorithm with various choices of
parameters, for the Metropolis reflection and for the heat bath
algorithm.  The simulations were done at $\beta=2.2$, on an 8 by 8
lattice.  We always performed one million full lattice sweeps.  In the
Metropolis cases we also quote acceptance rates.  In all cases we
measured the CPU time consumed for 20,000 sweeps on a
Pentium 166 MMX PC.  The last column contains a number $r_{\rm eff}$
which is obtained by multiplying the autocorrelation time by the CPU
factor, and then normalizing such that $r_{\rm eff}$ is one for the
heat bath algorithm.

\begin{table}
\small 
\begin{center}
  \begin{tabular}{|c||r|c|c|c|r|c|}
 \hline
algorithm & $\epsilon$ \phantom{x} & $n_{\rm hit}$ & 
$\tau_{{\rm int},\chi_{\II}}$ \phantom{xx} & acc & cpu & $r_{\rm eff}$ \\
\hline 
M &  1.5\phantom{0}  & 1  & 43.00 $\pm$ 1.40  & 47\% &  7.1 &  0.943(31) \\
M &  2.5\phantom{0}  & 1  & 34.40 $\pm$ 1.00  & 30\% &  7.3 &  0.776(23) \\
M &  3.0\phantom{0}  & 1  & 35.92 $\pm$ 1.04  & 26\% &  7.4 &  0.821(24) \\
M &  3.0\phantom{0}  & 2  & 20.40 $\pm$ 0.44  & 51\% & 14.2 &  0.895(19) \\
M &  3.14 & 1  & 36.00 $\pm$ 1.08  & 25\% &  7.4 &  0.823(25) \\
M &  3.14 & 2  & 21.56 $\pm$ 0.48  & 48\% & 14.5 &  0.966(22) \\
MR&       &    & 29.68 $\pm$ 0.56  & 25\% &  4.1 &  0.376(07) \\
HB&       &    & 11.16 $\pm$ 0.20  &      & 29.0 &  1.000     \\
\hline
  \end{tabular}
\parbox[t]{.85\textwidth}
{
\caption{\label{compL8}
\small
  Integrated autocorrelation times of $\chi_{\II}$, $L=8$.
  acc denotes the acceptance rate. cpu is a rough
  estimate of CPU, in seconds, for 20,000 sweeps.
  $r_{\rm eff}$ is obtained from multiplying the
  integrated autocorrelation time by the cpu factor and normalizing
  such that it is one for the heat bath algorithm.
}}
\end{center}
\end{table}

The heat bath algorithm has by far the smallest autocorrelation times.
However, the Metropolis reflection algorithm implementation is so much
faster that it effectively becomes the most efficient algorithm among
the ones investigated. In the combination with the OR algorithm to be
discussed in the next section we always used two subsequent Metropolis
reflection algorithm sweeps as the basic unit.  Let us remark that
using different algorithm implementations or computers
we could have reached different conclusions.

We now investigated the critical slowing down of the Metropolis
reflection algorithm by running it on various lattice sizes, always at
$\beta=2.2$.
\begin{table}
\small 
\begin{center}
  \begin{tabular}{|r||c|c|r|}
    \hline
    $L$ & $\tau_{{\rm int},E_{\XY}}/L^{2.3}$ &
          $\tau_{{\rm int},\chi_{\II}}/L^{2.3}$ 
      & S/$10^6$ \\
    \hline
   8 &    0.1051(17) &   0.1306(21) &  1.3\phantom{x} \\
  12 &    0.0763(17) &   0.1082(28) &  1.3\phantom{x} \\
  16 &    0.0760(17) &   0.1092(27) &  2.5\phantom{x} \\
  20 &    0.0753(17) &   0.1050(27) &  3.8\phantom{x} \\
  24 &    0.0813(17) &   0.1112(28) &  6.3\phantom{x} \\
  32 &    0.0825(17) &   0.1074(31) &  9.0\phantom{x} \\
  40 &    0.0798(17) &   0.1080(38) & 10.0\phantom{x} \\
  48 &    0.0796(17) &   0.1087(43) & 12.0\phantom{x} \\
  64 &    0.0871(17) &   0.1117(60) & 13.7\phantom{x} \\             
    \hline
  \end{tabular}
\parbox[t]{.85\textwidth}
 {
\caption{\label{CSDmetro}
\small 
Integrated autocorrelation times of $E_{\XY}$ and 
$\chi_{\II}$ for the Metropolis reflection algorithm, rescaled
by a factor $L^{2.3}$ in both cases. 
The correlation times are measured in units of two sweeps. 
$S$ denotes the statistics in the same units. 
}}
\end{center}
\end{table}
Table \ref{CSDmetro} shows our findings for lattice sizes up to 
$L=64$. The integrated autocorrelation times have been rescaled
by a factor $L^{2.3}$. The $\tau_{\rm int}$ become really large. 
For $L=64$ we observe 
$\tau_{{\rm int},\chi_{\II}}= 1593 \pm 86$. 
Fits with the law 
\be\label{fitlaw}
\tau_{\rm int} = A \cdot L^{z} 
\ee
yield 

\begin{center}
\begin{tabular}{clcc}
            &  \phantom{x}  $A$      &  $z$     &  $\chi^2$/dof  \\
\hline  
$E_{\XY}$  &  0.064(4)    &  2.36(2) &  1.4 \\ 
$\chi_{\II}$& 0.106(8)   &  2.31(2) &  0.5 \\
\hline 
\multicolumn{4}{c}{ } \\[-2mm]
\multicolumn{4}{c}{\small \bf 2 Metropolis Reflections} \\
\end{tabular}
\end{center}

\vspace{5mm}

In both cases the data from $L=8$ were discarded from the fit. 
Our error bars take into account statistical errors only. 

\section{Efficiency of Overrelaxation Steps}

We decided to examine a combination of two Metropolis
reflection sweeps with $N$ OR sweeps. It is a general
observation that $N$ should scale like the correlation length or, if
the system is at criticality, with the lattice size $L$. The latter
case is relevant for us. 

In order to find the right mixture of reflection and OR steps, we
computed $\tau_{\rm int}$ for the XY energy and the Ising
susceptibility.  The runs were done on SUN sparc stations, for lattice
sizes 8, 16, and 32. Our results for the two smaller lattices are
quoted in tables \ref{tintL8} and \ref{tintL16}.  In the particular
implementation of our code, the overrelaxation sweeps needed $\approx$
1/16 of the run time of two MR sweeps. This CPU factor enters
into the numbers $\tau_{\rm int}^{\rm corr}$ defined through
\be
\tau_{\rm int}^{\rm corr} = \tau_{\rm int} \, \left(1 + \frac{N}{16}
\right) \, .  
\ee 
The behavior of $\tau_{\rm int}^{\rm corr}$ in both
tables together with the $L=32$ results suggest that it is reasonable to mix
according to $N = \frac{3}{8} \, L \,$. Of course this would have 
come out differently for other CPU time ratios between the MR and the
OR parts of the update procedure. On a Pentium PC, we measured 
a factor of 1/6 between the two CPU times. 

\begin{table}
\small 
\begin{center}
  \begin{tabular}{|r||r@{$\pm$}l|r@{$\pm$}l||r@{$\pm$}l|r@{$\pm$}l|}
    \hline
    $N$ & \multicolumn{2}{c|}{$\tau_{{\rm int},E_{\XY}}$} &
          \multicolumn{2}{c||}{$\tau_{{\rm int},E_{\XY}}^{\rm corr}$} &
          \multicolumn{2}{c|}{$\tau_{{\rm int},\chi_{\II}}$} &
          \multicolumn{2}{c|}{$\tau_{{\rm int},\chi_{\II}}^{\rm corr}$} \\
    \hline
      0 &  9.54 & 0.26 &  9.54 & 0.26  & 11.88 & 0.41 & 11.88 & 0.41 \\
      1 &  5.82 & 0.13 &  6.18 & 0.14  &  5.63 & 0.12 &  5.98 & 0.13 \\
      2 &  5.50 & 0.12 &  6.19 & 0.14  &  4.84 & 0.09 &  5.45 & 0.10 \\
      3 &  5.39 & 0.11 &  6.40 & 0.13  &  4.50 & 0.08 &  5.34 & 0.10 \\
      4 &  5.39 & 0.11 &  6.74 & 0.14  &  4.41 & 0.08 &  5.51 & 0.10 \\
      5 &  5.18 & 0.11 &  6.80 & 0.14  &  4.21 & 0.07 &  5.53 & 0.09 \\
      6 &  4.88 & 0.09 &  6.71 & 0.12  &  3.94 & 0.08 &  5.42 & 0.11 \\
      7 &  4.79 & 0.09 &  6.89 & 0.13  &  3.81 & 0.06 &  5.48 & 0.09 \\
      8 &  4.76 & 0.09 &  7.14 & 0.14  &  3.68 & 0.06 &  5.52 & 0.09 \\
    \hline
  \end{tabular}
\parbox[t]{.85\textwidth}
 {
\caption{\label{tintL8}
\small 
Integrated autocorrelation times for $E_{\XY}$ and 
$\chi_{\II}$, $L=8$. Two MR sweeps are blended 
with $N$  OR sweeps,
$\beta=2.2$. 
}}
\end{center}
\end{table}

\begin{table}
\small 
\begin{center}
  \begin{tabular}{|r||r@{$\pm$}l|r@{$\pm$}l||r@{$\pm$}l|r@{$\pm$}l|}
    \hline
    $N$ & \multicolumn{2}{c|}{$\tau_{{\rm int},E_{\XY}}$} &
          \multicolumn{2}{c||}{$\tau_{{\rm int},E_{\XY}}^{\rm corr}$} &
          \multicolumn{2}{c|}{$\tau_{{\rm int},\chi_{\II}}$} &
          \multicolumn{2}{c|}{$\tau_{{\rm int},\chi_{\II}}^{\rm corr}$} \\
    \hline
      0 & 32.70 & 1.87 &  32.70 & 1.87  & 56.94 & 3.95 &  56.94 & 3.95 \\
      1 & 13.25 & 0.43 &  14.08 & 0.46  & 20.29 & 0.82 &  21.56 & 0.87 \\
      2 & 10.57 & 0.31 &  11.89 & 0.35  & 13.89 & 0.46 &  15.63 & 0.52 \\
      3 &  9.60 & 0.26 &  11.40 & 0.31  & 10.85 & 0.32 &  12.88 & 0.38 \\
      4 &  9.08 & 0.26 &  11.35 & 0.33  & 10.24 & 0.30 &  12.80 & 0.38 \\
      5 &  8.60 & 0.23 &  11.29 & 0.30  &  8.96 & 0.24 &  11.76 & 0.32 \\
      6 &  7.79 & 0.19 &  10.71 & 0.26  &  7.87 & 0.20 &  10.82 & 0.28 \\
      7 &  7.72 & 0.23 &  11.19 & 0.33  &  7.61 & 0.21 &  10.94 & 0.30 \\
      8 &  7.67 & 0.21 &  11.51 & 0.32  &  7.38 & 0.17 &  11.07 & 0.26 \\
      9 &  7.69 & 0.19 &  12.02 & 0.30  &  7.20 & 0.17 &  11.25 & 0.27 \\
    \hline
  \end{tabular}
\parbox[t]{.85\textwidth}
 {
\caption{\label{tintL16}
\small 
Same as table \ref{tintL8}, but for $L=16$.
}}
\end{center}
\end{table}

In order to study the volume dependence of the $\tau_{\rm int}$ we 
made $ \geq 500,000$ update cycles on various lattice sizes, ranging from 
$L=8$ to $L=128$. A cycle consisted of two MR sweeps, 
followed by $3L/8$ OR sweeps. Our results for the $\tau_{\rm int}$ 
are quoted in table~\ref{CSDor}. 

\begin{table}
\small 
\begin{center}
  \begin{tabular}{|r||c|c|c|}
    \hline
    $L$ & $\tau_{{\rm int},E_{\XY}}/L^{1.13}$ &
          $\tau_{{\rm int},\chi_{\II}}/L^{1.35}$ 
      & S/$10^6$ \\
    \hline
   8  &    0.514(10) &   0.2717(48) &   0.5 \\
  12  &    0.430(10) &   0.2343(56) &   0.5 \\
  16  &    0.340(08) &   0.1864(47) &   0.5 \\
  20  &    0.341(10) &   0.1926(58) &   0.5 \\
  24  &    0.322(10) &   0.1822(60) &   0.5 \\
  32  &    0.289(12) &   0.1656(62) &   0.5 \\
  40  &    0.296(12) &   0.1694(76) &   0.5 \\
  48  &    0.290(13) &   0.1714(86) &   0.5 \\
  64  &    0.262(14) &   0.1564(98) &   0.5 \\
  80  &    0.291(08) &   0.1720(57) &   2.5 \\
  96  &    0.297(09) &   0.1727(62) &   2.5 \\
 128  &    0.288(10) &   0.1606(69) &   1.5 \\
    \hline
  \end{tabular}
\parbox[t]{.85\textwidth}
 {
\caption{\label{CSDor}
\small 
Integrated autocorrelation times for $E_{\XY}$ and 
$\chi_{\II}$ of the Metropolis reflection algorithm, 
blended with $N = 3L/8$ OR sweeps. The autocorrelation 
times have been rescaled by factors $L^{1.13}$ and 
$L^{1.35}$, respectively. 
A unit of measurement is two Metropolis reflection sweeps combined with 
3$L$/8 OR sweeps.
$S$ denotes the statistics in these units. 
}}
\end{center}
\end{table}
Fitting with the law eq.~(\ref{fitlaw}), discarding lattice sizes 
$L \le 24$, we obtained 

\begin{center}
\begin{tabular}{clcc}
            &  \phantom{x}  $A$      &  $z$     &  $\chi^2$/dof  \\
\hline  
$E_{\XY}$   & 0.28(5) &  1.13(4) &  1.0 \\ 
$\chi_{\II}$& 0.17(3) &  1.34(4) &  0.8 \\
\hline 
\multicolumn{4}{c}{ } \\[-2mm]
\multicolumn{4}{c}{\small \bf 2 Metropolis Reflections + 3$L$/8 OR} \\
\end{tabular}
\end{center}

\vspace{5mm}

For a fair comparison with the Metropolis reflection algorithm one
needs to use units that take into account the work spent on the OR
steps.  In fair work units the scaling law is
\be 
\tau_{\rm int} = A \cdot L^{z} \, \left( 1 + \omega \, 3 L /8 \right) \, , 
\ee 
with
\be
\omega = \frac12 \, \frac{\mbox{CPU cost of OR sweep}}
{\mbox{CPU cost of MR sweep}} \, . 
\ee
The asymptotic behaviour will thus be governed by an exponent $z+1$, 
with an amplitude $ 3/8 \, \omega \, A$. Comparing the results 
of the OR and the pure Metropolis algorithms, we find compatible 
or similar exponents. The effective amplitudes, however, differ. 
In case of the chiral susceptibility, the net efficiency gain 
for large lattices is 
\be 
\gamma = \frac38 \; \omega \; \frac{0.17(3)}{0.106(8)} \, . 
\ee 
For realistic values of $\omega$ between 0.1 and 0.2 this gives 
a speedup of about 10. This statement is in good agreement 
with the actual values of autocorrelations times observed 
in our simulations. 

It is interesting to look also at the performance of an 
algorithm with a fixed admixture of OR steps. We made simulations 
with $N=3$ kept fixed for all lattice sizes. Our results are 
summarized in table~\ref{CSDor3}. 

\begin{table}
\small 
\begin{center}
  \begin{tabular}{|r||c|c|c|}
    \hline
    $L$ & $\tau_{{\rm int},E_{\XY}}/L^{2.32}$ &
          $\tau_{{\rm int},\chi_{\II}}/L^{2.20}$ 
      & S/$10^6$ \\
    \hline
   8 &   0.04169(96) &   0.04391(93) &   0.3 \\
  16 &   0.01679(39) &   0.02867(72) &   0.5 \\
  20 &   0.01495(24) &   0.02788(55) &   1.3 \\
  24 &   0.01419(19) &   0.02722(46) &   2.5 \\
  32 &   0.01066(19) &   0.02280(49) &   2.5 \\
  36 &   0.00988(20) &   0.02235(53) &   2.5 \\
  40 &   0.01004(23) &   0.02388(66) &   2.5 \\
  48 &   0.01029(20) &   0.02385(56) &   5.0 \\
  64 &   0.00945(35) &   0.02080(90) &   2.5 \\
  72 &   0.01048(27) &   0.02362(70) &   8.0 \\
  96 &   0.00961(42) &   0.02315(18) &   5.0 \\
    \hline
  \end{tabular}
\parbox[t]{.85\textwidth}
 {
\caption{\label{CSDor3}
\small 
Integrated autocorrelation times for $E_{\XY}$ and 
$\chi_{\II}$ for the Metropolis reflection algorithm, 
blended with $N = 3$ OR sweeps. The autocorrelation 
times have been rescaled by factors $L^{2.23}$ and 
$L^{2.20}$, respectively. 
A unit of measurement is two Metropolis reflection steps combined with 
3 OR steps.
$S$ denotes the statistics in these units. 
}}
\end{center}
\end{table}

Fitting with the scaling law eq.~(\ref{fitlaw}), we had to discard 
data from $L \leq 32$. The results are 

\begin{center}
\begin{tabular}{clcc}
            &  \phantom{x}  $A$      &  $z$     &  $\chi^2$/dof  \\
\hline  
$E_{\XY}$   & 0.10(1)  &  2.33(3) &  2.2 \\ 
$\chi_{\II}$& 0.022(3) &  2.21(3) &  2.5 \\
\hline 
\multicolumn{4}{c}{ } \\[-2mm]
\multicolumn{4}{c}{\small \bf 2 Metropolis Reflections + 3 OR} \\
\end{tabular}
\end{center}

\vspace{5mm}

Assuming an  $\omega$ about 0.15, the net efficiency gain over 
the Metropolis reflection without OR steps 
is of order 3 (for $\chi_{\II}$).

\section*{Conclusion}
The net efficiency gain of an inclusion of OR steps in a Monte Carlo
procedure for the FFXY model might vary with implementation and fine
tuning details. It does not seem to improve on the dynamical critical
exponents in a significant way. However, for a given lattice size, it
definitely enhances the efficiency of the updating.  The fact that it
can be implemented so easily makes its usage mandatory.  It would be
interesting to combine the OR updating with multilevel techniques.
E.g., one could perform simultaneous rotations of all spins in a block
of size $l$ by $l$ such that the total energy remains unchanged.


\begin{thebibliography}{99}

\bibitem{review} H.S.J. van der Zant et al., Physica B 152 (1988) 56. 

\bibitem{denniston} C. Denniston and C. Tang, Phys.\ Rev.\ Lett.\ 75 
                    (1995) 3930.  

\bibitem{ling} X.S. Ling et al., Phys.\ Rev.\ Lett.\ 76 (1996) 2989.

\bibitem{KT}   J.M. Kosterlitz and D.J. Thouless, 
               J. Phys.\ C 6 (1973) 1181.

\bibitem{Teitel} S. Teitel and C. Jayaprakash, 
                 Phys. Rev. B 27 (1983) 598. 

\bibitem{Berge} B. Berg\'e, H.T. Diep, A. Ghazali, and  
                P. Lallemand, Phys.\ Rev.\ B 34 (1986) 3177. 

\bibitem{Thijssen} J.M. Thijssen and H.J.F. Knops, 
                   Phys.\ Rev.\ B 37 (1988) 7738. 

\bibitem{Scheinine} A.L. Scheinine, Phys.\ Rev.\ B 39 (1989) 9368. 

\bibitem{Nicolaides} D.B. Nicolaides, J. Phys.\ A 24 (1991) L231. 

\bibitem{Knops} Y.M.M. Knops, B. Nienhuis, H.J.F. Knops, 
                and  H.W.J. Bl\"ote, Phys.\ Rev.\ B 50 (1994) 1061. 

\bibitem{Lee}    J.-R. Lee, Phys. Rev. B 49 (1994) 3317.

\bibitem{LeeLee} S. Lee and  K.-C. Lee, 
                 Phys.\ Rev.\ B 49 (1994) 15184. 

\bibitem{RSJ}    G. Ramirez-Santíago and J.V. Jos\'e, 
                 Phys. Rev. B 49 (1994) 9567. 

\bibitem{XYIsing} M.P. Nightingale, E. Granato, and J.M. Kosterlitz, 
                  Phys. Rev. B 52 (1995) 7402.  

\bibitem{Olsson1} P. Olsson, Phys.\ Rev.\ Lett.\ 75 (1995) 2758. 

\bibitem{Olsson2} P. Olsson, Phys.\ Rev.\ B 55 (1997) 3585.

\bibitem{SW}  R.H. Swendsen and J.S. Wang, 
              Phys.\ Rev.\ Lett.\ 58 (1987) 86.

\bibitem{Wolff} U. Wolff, Nucl.\ Phys.\ B 322 (1989) 759,
                U. Wolff, Phys.\ Rev.\ Lett.\ 62 (1989) 361.

\bibitem{KanDom} D. Kandel and E. Domany, 
                 Phys.\ Rev.\ B 43 (1991) 8539.

\bibitem{Coddington} P.D. Coddington and L. Han, 
                     Phys.\ Rev.\ B 50 (1995) 3058. 

\bibitem{KBD}   D. Kandel, R. Ben-Av, and E. Domany, 
                Phys.\ Rev.\ B 45 (1992) 4700. 

\bibitem{Adler}     S.L. Adler, Phys.\ Rev.\ D 23 (1981) 2901. 

\bibitem{Creutz} M. Creutz, Phys.\ Rev.\ D 36 (1987) 515.    

\bibitem{BrownWoch} F. Brown and T.J. Woch, Phys.\ Rev.\ Lett 58 (1987)  2394.

\bibitem{Aposto} J. Apostolakis, C.F. Baillie, and G.C. Fox, 
                 Phys.\ Rev.\ D 43 (1991) 2687. 

\bibitem{BaiGup} C.F. Baillie and R. Gupta, Phys.\ Rev.\ B 45 (1992) 2883. 

\bibitem{CP3} U. Wolff, Phys.\ Lett.\ B 284 (1992) 94. 

\bibitem{ORWolff}   U. Wolff, Phys. Lett. B 288 (1992) 166.

\bibitem{HorKen}    I. Horváth and A.D. Kennedy, 
                    Nucl.\ Phys.\ B 510 (1998) 367. 

\bibitem{Sokal} A.D. Sokal, {\em Monte Carlo Methods in Statistical Mechanics:
                Foundations and New Algorithms}, Cours de Troisi\`eme 
                Cycle de la Physique en Suisse Romande, Lausanne, 1989. 

\bibitem{hattori} T. Hattori and H. Nakajima, 
                  Nucl.\ Phys.\ Proc.\ Suppl.\ 26 (1992) 635. 

\end{thebibliography}
\end{document}